# Multivariate Residual Estimation Risk

D.J. Manuge
*Chief Technology Officer*
*Corl Financial Technologies Inc.*

## 1   Introduction

Financial institutions have become increasingly reliant on internal models to estimate valuation or risk parameters such as profit-and-loss (P&L), probability of default (PD), loss given default (LGD), or exposure at default (EAD). Prior to model approval and implementation in a production environment, these models must pass rigorous validation assessments to ensure they are compliant with internal and external regulatory requirements. One of the crucial objectives driving the validation process is the measurement and management of model risk [18]. The concern of model risk is well motivated, as financial institutions rely heavily on quantitative analysis and models in many aspects of their financial decision-making process. To address this risk, a core element of the validation process is outcome analysis, specifically back-testing, which is the comparison of model estimates to realized values [1].

At a high level, back-testing is conducted to assure oneself (and others) that a model produces reasonable estimates of some target variable. One potential question it aims to answer is: "how much buffer should I add to my estimates to cover the uncertainty with estimation?" This question arises often in the context of capital adequacy, where a deposit-taking institution must hold a certain amount in reserves to establish a reasonable margin of safety, or where a firm must conservatively estimate the Value-at-Risk of its trading book to ensure hypothetical profit-and-loss falls below this value at a sufficiently low frequency. Fortunately, this question has recently been formalized via the concept of *residual estimation risk.*

To better understand residual estimation risk, consider a firm that attempts to estimate the loss of capital in one of their portfolios. This quantity may be known as the *capital estimator*. The firm acknowledges that the capital estimator is a mere estimation, and in lieu of this, would like to determine the inherent risk in relying on such an estimate. This inherent risk can be viewed as the quantity that remains after the portfolio incurs realized losses. This residual amount is specifically the quantity that the firm aims to measure; the residual estimation risk (herein referred to as RER). Initially introduced by Bignozzi and Tsanakas in [1, 2, 3], the authors define RER as any risk measure applied to the difference between actual losses and a capital estimator. They provide thorough results on how to estimate RER under parametric and empirical loss distributions, eliminate RER by adjusting the capital estimator via bootstrapping, as well as how to explicitly calculate parameter uncertainty, model uncertainty, and model risk under various distributional assumptions.

For market risk models that estimate Value-at-Risk, this framework is natural. Practitioners are often presented with some historical loss distribution $X = (X_1, \ldots, X_n) \in R^n$ that is to be used to determine the Value-at-Risk $\eta(X) \in R$. However, the concept of quantifying the excess risk in an estimate is not specific to loss distributions or even market risk models. Any model that produces risk estimates for use by financial institutions often requires valid back-testing against historical or simulated data, and RER is capable of measuring this error. The existing definition of RER is available for such quantification, provided that the estimator is a function $\eta: R^n \to R$. However, this is not always the case. In credit risk, model outputs are often represented by a vector that describes the risk of individuals or segmented subsets of a population. For example, consider a bank issuing credit to a group of publicly traded companies. The bank may develop an LGD estimation model that segments publicly traded companies into distinguishable risk ratings; such as those defined by Moody's, S&P, or an internal ratings system. The realized (or actual) LGDs of the companies are used to calibrate the model, and ultimately used to classify each company into a distinct yet homogenous risk segment. In this framework, each segment corresponds to a quantifiable LGD that is inherited by members of that segment. Effectively, through this process each publicly traded company is assigned an estimate for LGD. Mathematically, this model describes a surjective function between actual LGDs and estimated LGDs; the condition being that $\eta: R^n \to R^m$ where $m \leq n$. In the case where $m = 1$, RER is quantified under the originally-motivated definition. However, when a set or vector of estimates is the output of a model, a multivariate definition is required.

As such, the purpose of this paper is to describe and extend the use of the newly-introduced measure, *residual estimation risk*. Following the seminal work of Bignozzi and Tsanakas, the quantification of residual estimation risk is proposed in a multivariate framework. Our aim is to provide a succinct and practical introduction to the concept, to motivate its use as a back-testing measure, and to provide examples related to credit risk parameter estimation. In section 2, we introduce RER defined by various risk measures, and illustrate the calculation using R and SAS®. In section 3, we propose a back-testing criterion for the measure, which can be altered to assess model performance for both accuracy and conservatism. In section 4, we conduct back-testing on risk parameter estimates of retail credit portfolios, including multiple back-testing measures for comparison. Finally, we conclude our findings and propose areas for future work in section 5.

## 2   Residual Estimation Risk

Suppose $Y \in \mathbb{R}^n$ is a random variable defined on some probability space such that $Y \sim F$. Let a risk measure $\rho$ be a functional Y satisfying the following properties:

i. Monotonicity: If $Y_1 \leq Y_2$ then, $\rho(Y_1) \leq \rho(Y_2)$.
ii. Translation Invariance: If $a \in \mathbb{R}$, $\rho(Y + ae) = \rho(Y) + a$.
iii. Positive Homogeneity: If $\lambda \geq 0$, $\rho(\lambda Y) = \lambda \rho(Y)$.
iv. Law Invariance: If $Y_1 \equiv Y_2$, $\rho(Y_1) = \rho(Y_2)$.[1]

---

[1] Where $\equiv$ denotes distributional equality.

Some common risk measures that satisfy the above properties are Value-At-Risk (VAR), Expected Shortfall (ES)[2], and Restricted Value-At-Risk (RVAR).

The distribution of Y is often not known in practice, and requires estimation using empirical data $X = (X_1, \ldots, X_m)$. However, if X and Y are independent and X,Y ~ F, then $\rho(X) = \rho(Y)$. In other words, the risk measure of Y can be estimated using X by means of an estimator $\eta(X)$ such that $\rho(\eta(X)) \cong \rho(Y)$. In practice, an estimator can be viewed as a model that produces estimates of Y using data from X. The difference between a value sampled from Y and the estimator $\eta(X)$ can be represented as a random variable in $\mathbb{R}^n$. The difference between multiple samples forms an error distribution that illustrates the error in adopting the estimator. This error can be measured via residual estimation risk per the definition;

$$RER := \rho(Y - \eta(X)). \quad (1)$$

Equivalently, $RER$ is the additional amount that should be added to the current estimate $\eta(X)$ to accurately reflect outcomes distributed from $Y$. This statement is mathematically equivalent to

$$\rho(Y - (\eta(X) + RERe)) = 0. \quad (2)$$

Where $e = (1, \ldots, 1)^T$. In this view, RER can be considered the constant buffer that should be added to model estimates to eliminate the error from estimation. Clearly, RER is sensitive to the shape of the error distribution and selection of risk measure. From (2), one can infer the following. If RER is negative, the model estimates are sufficient in covering actuals that may result from the true distribution. In this case, one may conclude that the model used to generate $\eta(X)$ produces conservative enough estimates. Conversely, if RER is positive, the model estimates are not sufficient in covering exposures that may result from the true distribution. In this case, we may conclude that the model does not produce conservative enough estimates of $\hat{Y}$. Since the distribution of $\hat{Y}$ underestimates the risk of the true distribution, it should be interpreted as a warning signal. For if the historical distribution is any indication of the future, use of these estimates may continuously underestimate the portfolio (at some pre-specified level of confidence). Finally, when RER is zero we have completely eliminated any excess risk; i.e. the model estimates exactly cover the actuals that may result from the true distribution. It can be said that when $RER = 0$, the resulting model estimates or estimator is *optimal.* Indeed, there exists an estimator such that all model estimates are optimal regardless of the risk measure, level of confidence, or underlying distributions.

---

[2] This is also known as Tail Value-at-Risk.

**Lemma**: The estimator $\eta(X) = \rho(X)e$ is optimal.

Proof: Let $\eta(X) := \rho(X)e$. Then, by the properties of translation and law invariance, $RER = \rho(Y - \eta(X)) = \rho(Y - \rho(X)e) = \rho(Y) - p(X) = \rho(Y) - \rho(Y) = 0$.

Of course, this optimal estimator is not unique. There are many models capable of producing estimates that eliminate estimation risk. In fact, most models used in the quantification of risk are indeed optimal for some risk measure and level of confidence. Suppose we've developed a model that produces estimates given some underlying distribution. One may pose the question; at what level of confidence are these estimates optimal? To find such a confidence level, we must solve for the confidence level such that $RER = 0$. Formulating the following optimization problem provides us with the answer:

$$\tilde{p} = \arg\min_{0 \leq p \leq 1} |\rho(Y - \eta(X))|$$

From a computational perspective, some limiting factors may deter us from obtaining a perfect elimination of RER, such as the size of data and the shape of the absolute error distribution. When computational limitations arise, we converge to the optimal $p$ such that RER is *closest* to zero. Often, these limiting factors only affect the results in extreme cases, and a large data size almost always has the ability to offset any precision lost from other limiting factors.

When solving the inverse problem, one may want to know what an acceptable value is for $\tilde{p}$. This value is sensitive to the error distribution, risk measure, and depends on the level of conservatism one aims to achieve. Hence, there is no universal threshold. As alluded to prior, for economic capital models one might relax the inquisition for conservatism and consider a lower optimal confidence level as acceptable. However, for stress testing models one may seek a higher level of conservatism, and thus a higher optimal confidence level. Some guiding principles can be established knowing the risk measure used for RER.

## 2.1 Risk Measures for Residual Estimation Risk

Since the argument of RER is the absolute error[3] of the model outputs and actual outcomes, the choice of measure will have a profound impact on the results obtained. To better understand what RER is measuring, suppose we have a model that estimates the probability of default $\hat{Y}$ of a credit portfolio. We wish to investigate RER under various risk measures; in particular Value-at-Risk and Expected Shortfall.

### *2.1.1 Value-at-Risk*

Let our risk measure be Value-At-Risk (VAR), defined as

$$\rho(X) \coloneqq VaR_p(X) = \inf\{m \in R \mid P(X \leq m) \geq p\} \qquad (3)$$

---

[3] We define the absolute error as $Y - \hat{Y}$.

on some distribution of $X = Y - \hat{Y}$. Then $\rho(X)$ represents the maximum error the estimated credit portfolio is expected to attain within $(1 - p) \cdot 100\%$ confidence. For example, by setting $p = 0.05$ we would expect to achieve an error as high as $\rho(X)$ about once in every twenty observations. This raises the question; provided we adopt the model that produces these estimates, would the estimates sufficiently cover observations resulting from the known portfolio? RER measures exactly that. It is the amount of excess risk as a result of adopting a model to estimate $\hat{Y}$ when the true distribution $Y$ is known. Since VAR is a median-based measure, this excess risk is concerned with the frequency of observations that are underestimated. When solving for the optimal confidence interval (i.e. inverse problem), it is immediately obvious that;

i. If $\tilde{p} < 0.5$, the estimated risk parameters overestimate the true risk parameters for *less* than half of the portfolio.
ii. If $\tilde{p} > 0.5$, the estimated risk parameters overestimate the true risk parameters for *more* than half of the portfolio.

This enables us to define a pivot point for RER under VAR, where the numbers of accounts that are over and underestimated are equal.

For an arbitrary distribution we can establish an upper bound on RER under VAR. This result follows from Chebyshev's Inequality which assigns an upper bound on the number of extreme observations that can exist in an arbitrary distribution. Mathematically, if $X$ is a random variable with finite mean $\mu_X$ and finite non-zero variance $\sigma_X^2$, then for positive $k \in R$,

$$\Pr(|X - \mu_X| \geq k\sigma_X) \leq \frac{1}{k^2}$$

This places an upper bound on the probability of observing values above $k\sigma$ and below $-k\sigma$.[4] Equivalently, this provides us with the range of observations that are expected to occur within k standard deviations of the mean. Following from the above inequality,

$$\frac{1}{k^2} \geq \Pr(|\mu_X - X| \geq k\sigma_X) = \Pr(\mu_X - X \geq k\sigma_X) \cup \Pr(\mu_X - X \leq -k\sigma_X) \geq \Pr(\mu_X - X \leq -k\sigma_X)$$

$$= \Pr(-X \leq -k\sigma_X - \mu_X) = \Pr(X \geq \mu_X + k\sigma_X) = 1 - \Pr(X \leq \mu_X + k\sigma_X)$$

Thus,

$$\frac{1}{k^2} \geq 1 - \Pr(X \leq \mu_X + k\sigma_X)$$

$$\Pr(X \leq \mu_X + k\sigma_X) \geq 1 - \frac{1}{k^2}$$

By setting $m = \mu_X + k\sigma_X$ and $p = 1 - \frac{1}{k^2}$ in the definition of $VaR$ from equation (3), and taking the infimum over $m$ of both sides we obtain

$$RER := VaR_{1-\frac{1}{k^2}}(X) \leq \mu_X + k\sigma_X.$$

This result places an upper bound on RER under VAR for any arbitrary error distribution where the mean and standard deviation are known.

If the error distribution is symmetric, we may obtain a sharper result.

$$\frac{1}{k^2} \geq \Pr(|\mu_X - X| \geq k\sigma_X) = \Pr(\mu_X - X \geq k\sigma_X) \cup \Pr(\mu_X - X \leq -k\sigma_X) = 2 \cdot \Pr(\mu_X - X \leq -k\sigma_X)$$

---

[4] For highly conservative models, one should expect a near-zero probability that model outputs fall into the left tail.

$$= 2 \cdot \Pr(-X \leq -k\sigma_X - \mu_X) = 2 \cdot \Pr(X \geq \mu_X + k\sigma_X) = 2 \cdot [1 - \Pr(X \leq \mu_X + k\sigma_X)]$$

Thus,

$$\frac{1}{k^2} \geq 2 - 2 \cdot \Pr(X \leq \mu_X + k\sigma_X)$$

$$\Pr(X \leq \mu_X + k\sigma_X) \geq 1 - \frac{1}{2k^2}$$

By setting $m = \mu_X + k\sigma_X$ and $p = 1 - \frac{1}{2k^2}$ in the definition of $VaR$ from equation (3), and taking the infimum over $m$ of both sides we obtain

$$RER := VaR_{1-\frac{1}{2k^2}}(X) \leq \mu_X + k\sigma_X.$$

This result places an upper bound on RER under VAR for any arbitrary symmetric error distribution where the mean and standard deviation are known. To demonstrate the improvement in sharpness, suppose $k = \sqrt{2}$. The upper bound for RER under VAR is equivalent in equation (2) and (3) for an arbitrary symmetric distribution. However, due to the monotonicity of the risk measure, it is obvious that $VaR_{0.5}(X) \leq VaR_{0.75}(X) \leq \mu_X + \sqrt{2}\sigma_X$. In other words, the confidence level of VAR improves from 50% to 75% using equation (3) when the error distribution is symmetric.

If we instead have the mean and standard deviation of the actual and estimated distributions, then we may decompose the upper bound into these components via

$$RER := VaR_{1-\frac{1}{k^2}}(Y - \hat{Y}) \leq \mu_Y - \mu_{\hat{Y}} + k\sqrt{\sigma_Y^2 + \sigma_{\hat{Y}}^2 - 2\,cov(Y, \hat{Y})}\ .$$

Since $\mu_X = E[X] = E[Y - \hat{Y}] = \mu_Y - \mu_{\hat{Y}}$ and $\sigma_X^2 = \sigma_{Y-\hat{Y}}^2 = \sigma_Y^2 + \sigma_{\hat{Y}}^2 - 2\,cov(Y,\hat{Y})$. The above inequality establishes an upper bound on RER under VAR for arbitrary distributions of $Y$ and $\hat{Y}$.[5] Since $VaR$ and the first two moments of the above distributions are known for any empirical distribution, one may solve for $k$ in this implicit equation to obtain the critical value where RER is completely eliminated. Setting RER=0 obtains

$$k^* \geq -\frac{\mu_X}{\sigma_X}$$

Since $\sigma_X > 0$ and $k > 0$, we have the following corollary: there exists a k* such that RER=0 if $\mu_X < 0$.

$$k^* \geq \frac{\mu_{\hat{Y}} - \mu_Y}{\sqrt{\sigma_Y^2 + \sigma_{\hat{Y}}^2 - 2\,cov(Y,\hat{Y})}}$$

Thus, we may explicitly solve for the number of standard deviations away from the mean of the error distribution $(X = Y - \hat{Y})$ that ensures RER is equal to zero.[6]

---

[5] We may further deduce that if $E[\hat{Y}]$ is an unbiased estimator of $E[Y]$, then $E\left[E[\hat{Y}] - E[Y]\right] = 0$, and since $k > 0$, we have that

$$RER \leq k\sqrt{\sigma_Y^2 + \sigma_{\hat{Y}}^2 - 2\,cov(Y,\hat{Y})} \leq 0.$$

In other words, RER is guaranteed to be non-positive when $E[\hat{Y}]$ is an unbiased estimator of $E[Y]$. This implies that $\hat{Y}$ produces conservative estimates of $Y$ for any confidence level.

[6] When $E[\hat{Y}]$ is an unbiased estimator of $E[Y]$, $RER \to 0$ when $k^* \to 0$.

The figures below demonstrate the sharpness of this upper bound for a normal distribution with respect to standard deviations away from the mean, k, and confidence level, 1-p. The plots indicate that as volatility of the distribution increases, as does the tightness of the bound. Furthermore, since the normal distribution has support on $-\infty \leq X \leq \infty$, as $k \to \infty$ or $k \to 1$, the error between RER and the upper bound approaches infinity. An equivalent relationship holds for when $p \to 0$ or $p \to 1$.

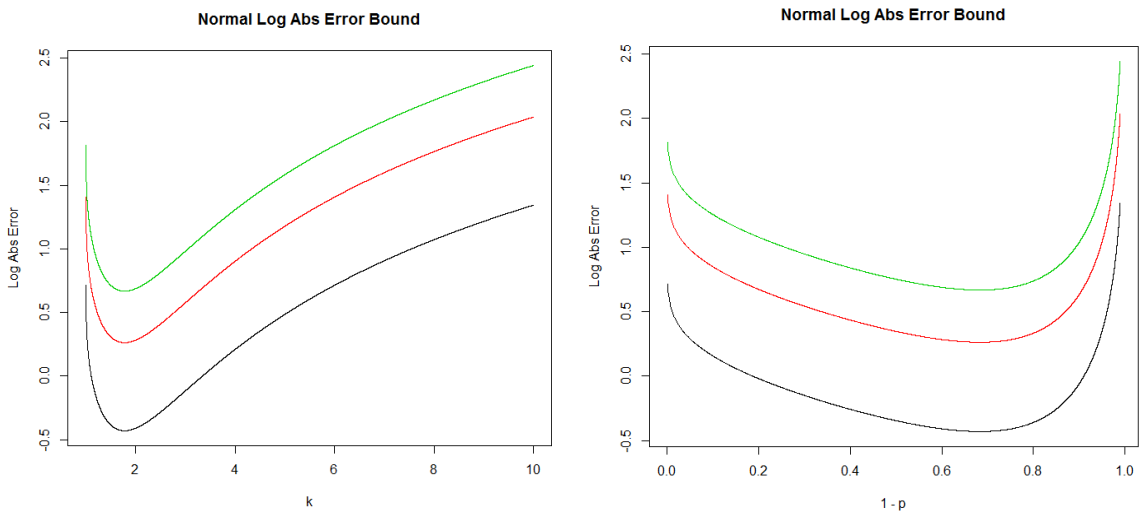

In the case where an arbitrary error distribution has support on $[A, B]$ where $-\infty \leq A, B \leq \infty$ we can deduce the following:

$$\lim_{k \to 1} \left| VaR_{1-\frac{1}{k^2}}(X) - \mu_X + k\sigma_X \right| = |VaR_0(X) - \mu_X + \sigma_X| = |A - \mu_X + \sigma_X|$$

Since

$$\lim_{k \to 1} VaR_{1-\frac{1}{k^2}}(X) = VaR_0(X) = A$$

and

$$\lim_{k \to \infty} \left| VaR_{1-\frac{1}{k^2}}(X) - \mu_X + k\sigma_X \right| = \infty$$

To compute RER under VAR, we may do so in SAS® via the function:

```
%MACRO VAR(dataset = , target = , prob = );

proc sort data = &dataset out = sorted_target;
by ⌖
run;

proc sql noprint;
select count(&target) into : number from sorted_target;
quit;

data sorted_target;
set sorted_target;
index = _N_;
run;

proc sql print;
select &target from sorted_target where index = floor(&number * &prob);
drop table sorted_target;
quit;

%MEND;
```

where *target* is the absolute error between the actuals and estimates stored in the dataset *dataset*.

### 2.1.2 Expected Shortfall

In a similar fashion, RER can be calculated using expected shortfall, defined as

$$\rho(X) \coloneqq \frac{1}{1-p} \int_p^1 VaR_\alpha(X)\, d\alpha = E[X|X > \inf\{m \in R \mid P(X \le m) \ge p\}]$$

where $0 \le p \le 1$ indicates the confidence level for ES. When $p = 0$, RER under ES reduces to the mean of the error. As such, there is no obvious pivot point for ES confidence levels (unlike VAR, where 0.5 corresponds to the fulcrum). Since ES measures the expectation of values that are above the VAR (for some $p$), one should expect optimal confidence levels to be much lower. However, for any risk measure, the higher the optimal confidence level, the more conservative the estimates. Since ES is arguably more delicate to the tails of the error distribution, ES will escalate if the magnitude of errors increase. This characteristic doesn't concern VAR because the measure is concerned with the number of underestimated observations, rather than the average of those underestimations.

Under ES, one may wish to determine what constitutes an acceptable level of confidence for the inverse problem. For example, if the optimal confidence level is above $p \cdot 100\%$, or equivalently RER is less than zero at a $p \cdot 100\%$ confidence level, then we may deduce the following;

i. The average error of $p \cdot 100\%$ of the most underestimated observations in the portfolio is less than zero.
ii. The average error after removing $(1-p) \cdot 100\%$ of the most overestimated observations in the portfolio is less than zero.

These statements can be interpreted as follows. Suppose we have a portfolio of mortgages for which we have estimated PD. It may be the case that the model tends to overestimate PD for a certain subgroup of the portfolio, and that this subgroup is related by certain variables that are not specified in the model. As an example, consider a geographic area whose economy depends primarily on a single industry (e.g. oil production and refining in areas of Alberta). While variables such as credit score, annual income, or days delinquent might be specified in the model, not all dependencies (regional or otherwise) might be captured in the model. If an event is triggered that negatively affects this subgroup via these unspecified variables, their realized default rates would increase, but the PD model may not adequately respond to this increase in risk. Under this scenario, previously overestimated mortgage PD would become less overestimated (perhaps underestimated), decreasing the conservatism of the model. If an event is triggered that removes these mortgages from the portfolio, a similar outcome would prevail. RER under ES measures this risk by quantifying the average error above a certain percentile of the error distribution. Since ES measures the expectation of values that are above the VAR (for some level $p$), one should expect confidence levels to be much lower for an arbitrary error distribution. In fact, *ceteris paribus,* by definition ES is always smaller than VAR.

To compute RER under ES, we may do so in SAS® via the function:

```
%MACRO ES(dataset = , target = , prob = );

proc sort data = &dataset out = sorted_target;
by ⌖
run;

proc sql noprint;
select count(&target) into : number from sorted_target;
quit;

data sorted_target;
set sorted_target;
index = _N_;
run;

proc sql print;
select mean(&target) from sorted_target where index > floor(&number * &prob);
drop table sorted_target;
quit;

%MEND;
```

where $target$ is the absolute error between the actuals and estimates stored in the dataset $dataset$. Regardless of the risk measure, the higher the optimal confidence level, the more conservative the estimates.

## 2.2   Empirical Distributions

Some measures of risk assume that the target variable follows a parametric distribution where parameters are estimated from a known sample. However, available samples may be small, noisy, or misrepresentative of the true distribution, leading to error in the estimated parameters. This parameter uncertainty has the potential to impact the output of statistical risk measures.[7] Furthermore, as is the nature of estimating natural phenomenon, the true distribution may not be known *a priori*. When this is the case, it is natural to assume that the empirical sample is representative of the true distribution. From a practical perspective this is almost always the case, and if often a prerequisite to developing a model or conducting a back-test. RER in this empirical distribution framework will be the aim of our discussion.

## 3 Model Back-Testing

To conduct back-testing, one must define what is meant by performance of the model in question. For example, in the context of stress testing models, preference may be towards producing highly conservative estimates, for economic capital models one may be partial towards producing lower conservative estimates, and for derivatives valuation models one may yearn for accuracy with market prices. The definition of model performance is indeed subject to discretion. While the choice of back-testing measure depends on the format or representation of the outputs, some commonly used measures in credit and market risk are the Accuracy Ratio, Mean-Squared Error, Akaike Information Criterion, Kolmogorov-Smirnov Statistic, Spearman Correlation, and Value-at-Risk – where each measure appropriately and uniquely quantifies different aspects of the model and its outputs. For this reason, it is essential that multiple measures are evaluated when back-testing to inform about the risk related to a wide range of model features and attributes.

---

[7] The influence of parameter uncertainty on risk measures has been studied in the context of insurance premium pricing (Mata), cost of claims for business lines (Borowicz and Norman), credit risk modelling (McNeil, Frey, and Embrechts), and banking revenue (Jorion).

Using the information specified above, the Basel II *traffic light approach* for market risk can be adapted as an RER back-testing criterion [14]. Suppose we obtain RER of a model at each point-in-time across a given time period. For the purpose of this back-test framework, when RER is positive at some point in time, a breach has occurred. A breach event under this definition implies under-conservatism of the estimates. The result of each point-in-time RER, determined by its sign, is recorded as a Bernoulli event (i.e. 1 and 0), where a breach corresponds to a success (1) and no breach corresponds to a failure (0). Since the result of each point-in-time RER is a Bernoulli event, the distribution of its sum follows a Binomial distribution. Hence, the observed sum of the breaches is compared against binomial probabilities to ensure the cumulative probability of observing this sum of breaches is within a reasonable expectation. In this case, the cumulative probability is the probability of obtaining a given number or fewer breaches across our given time period. The amount of trials, N, is generally defined by the frequency of the portfolio, where the number of trials is equal to the number of time-steps in the back-test period. In such a construct, the probability of observing a breach and the number of trials is known *apriori*. The probability of observing a breach, q, is equal to 1 minus the confidence level of the risk measure; and as previously mentioned, the confidence level of the risk measure will depend on the level of conservatism one aims to achieve.

*Conservatism* is broadly defined here to be the additional amount in model estimates relative to actual outcomes. The level of conservatism expected in a model should unsurprisingly differ depending on the model in question. Conservatism may be viewed as a compliance or back-testing objective, whose definition may be moulded by the purpose of the model being tested. For example, when calculating PD for economic capital, estimates might be low enough to be an accurate representation of the portfolio, yet on the side of conservatism to ensure underestimation is kept to a minimum. On the contrary, stressed PD estimates should be inherently inflated to provide a cushion during economic downturn periods. Once the objectives of the model are known, one may determine an appropriate confidence level for their back-test based on the risk measure. As an example, suppose VAR is the risk measure chosen to calculate RER.

To test lower levels of conservatism, we guide our selection using [1]. In reference to validating PD risk rating systems, Mui and Ozdemir motivate using a confidence level of 51% because;

i. The type I error (probability of rejecting the model when it is correct) is not more than 50%.
ii. Given i., the type II error (probability of accepting the model when it is not correct) is minimized.

In contrast, consider the case where a higher level of conservatism is desired. For VAR, the typical confidence levels of 90% or 95% may be appropriate choices, implying that 90% or 95% of the actual portfolio is overestimated by the model estimates. The choice of confidence level, as always, requires some judgment.

Once the parameters of the binomial distribution have been specified, the cumulative probabilities are used to split the back-test results into traffic light zones. Basel II defines the green zone as 0-95%, yellow zone as 95-99.99%, and red zone as 99.99%+ for the cumulative probability of breaches. As a result, these zones determine ranges for the number of observed breaches under the back-test. Traffic colours for a confidence level of 51% and 20 quarters are given by Table 3.1-1.

*Table 3.1-1: Traffic Light Zones for 20 quarters and 51% no breaches*

| # of Breaches: $X$ | $P(x < X)$ | Conclusion |
|---|---|---|
| $0 \leq X \leq 12$ | $0 \leq P(X) \leq 95\%$ | Acceptable |
| $13 \leq X \leq 17$ | $95\% < P(X) \leq 99.99\%$ | Monitoring |
| $18 \leq X \leq 20$ | $99.99\% < P(X) \leq 100\%$ | Enhancement |

This table is used to determine the conclusion of the back-test, and thus whether the estimates are sufficiently conservatism at a given confidence level. Note that it may be prudent to back-test at multiple confidence levels to eliminate any subjectivity arising from judgement.

## 4  Conclusions and Next Steps

In this manuscript, we introduce the concept of residual estimation risk (RER) as a measure of evaluating the conservatism inherent in adopting risk parameter estimates for a portfolio. RER can be stated as any risk measure applied to the difference between the true risk parameters and the estimated risk parameters.

In future work, one might consider calculating RER on the quarterly sub-portfolios. Here one can track RER over time to ensure that optimal levels of confidence and RER are consistently stable. Furthermore, it is well-known that bootstrapping an empirical distribution often leads to more accurate estimates of VAR. This procedure can be extended to RER, where we bootstrap on the true risk parameter distribution. For smaller data sizes, the improvement in accuracy from bootstrapping will be more pronounced.[8] While from a practical perspective we are only concerned with empirical distributions, exploration of RER with respect to parametric distributions may be a worthwhile endeavour. Furthermore, by placing restrictions on the risk measure, such as subadditivity or comonotone additivity, one might be able to determine the effect of diversification or obtain bounds on the size of RER. Finally, as an important verification tool, one might be concerned with the error in our estimation. This can be accomplished in various ways, such as placing confidence intervals on our risk measure, testing the failure rate, assessing the frequency of tail events (e.g. Kupiec test), or measuring the independence of tail events (e.g. Christoffersen test).

---

[8] Smaller data sizes might arise from having limited portfolio data, or when calculating segment-level/quarterly-level RER.